\documentclass[aps,prb,twocolumn,showpacs,superscriptaddress,longbibliography]{revtex4-2}  
\usepackage{graphicx}  
\usepackage{dcolumn}   
\usepackage{bm}        
\usepackage{amsmath}
\usepackage{amsthm}
\usepackage{amssymb}
\usepackage{bbold} 
\usepackage{mathrsfs}
\usepackage{array}
\usepackage[normalem]{ulem} 
\usepackage[dvipsnames,usenames]{color}
\usepackage{soul} 
\usepackage{physics}
\usepackage{soul}

\usepackage{color}
\usepackage[dvipsnames,usenames]{color}

\usepackage{braket}
\usepackage{verbatim}
\usepackage{float}
\usepackage{mhchem}
\usepackage[makeroom]{cancel}
\usepackage{upgreek}
\usepackage{commath}
\usepackage{siunitx}
\usepackage{orcidlink}
\usepackage{lipsum}

\usepackage[titletoc]{appendix}

\usepackage{hyperref}
\hypersetup{colorlinks=true, linkcolor=blue, citecolor=blue, urlcolor=blue}

\begin{document}
\title {Inferring stealthy hyperuniform correlations from quantum transport}

\author{Natanael C. Costa \orcidlink{0000-0003-4285-4672}}
\affiliation{Instituto de F\'isica, Universidade Federal do Rio de Janeiro, 21941-972 Rio de Janeiro -- RJ, Brazil}
\affiliation{Institut f\"ur Theoretische Physik und Astrophysik, Universit\"at W\"urzburg, 97074 W\"urzburg, Germany}
\author{Mauro S.~Ferreira \orcidlink{0000-0002-0856-9811}}
\affiliation{School of Physics, Trinity College Dublin, Dublin 2, Ireland}
\affiliation{Centre for Research on Adaptive Nanostructures and Nanodevices (CRANN) \& Advanced Materials and Bioengineering Research (AMBER) Centre, Trinity College Dublin, Dublin 2, Ireland} 
\author{Caio Lewenkopf~\orcidlink{0000-0002-2053-2798}}
\affiliation{Instituto de F\'isica, Universidade Federal do Rio de Janeiro, 21941-972 Rio de Janeiro -- RJ, Brazil}
\author{Felipe~A.~Pinheiro~\orcidlink{0000-0001-8712-0555}}
\affiliation{Instituto de F\'isica, Universidade Federal do Rio de Janeiro, 21941-972 Rio de Janeiro -- RJ, Brazil}
\author{Paul J. Steinhardt}
\affiliation{Department of Physics, Princeton University, Princeton, New Jersey, 08544, USA}
\author{Salvatore Torquato}
\affiliation{Department of Chemistry, Princeton University, Princeton, New Jersey 08544, USA}
\affiliation{Department of Physics, Princeton University, Princeton, New Jersey, 08544, USA}
\affiliation{Princeton Institute for the Science and Technology of Materials, Princeton University, Princeton, New Jersey 08544, USA}
\affiliation{Program in Applied and Computational Mathematics, Princeton University, Princeton, New Jersey 08544, USA}
\author{Carlo Vanoni\orcidlink{0000-0002-0927-2868}}
\affiliation{Department of Physics, Princeton University, Princeton, New Jersey, 08544, USA}


\begin{abstract}
Stealthy hyperuniform disordered systems exhibit strongly suppressed long-wavelength fluctuations, producing correlated disorder with unusual consequences for wave propagation. 
A central quantity characterizing these systems is the stealthiness parameter $\chi$, which controls the range of excluded Fourier components in the disorder spectrum. However, in realistic settings, the microscopic disorder configuration may not be directly accessible, making it challenging to determine $\chi$ from structural information alone. 
Here, we propose a conductance-based inverse protocol to recover stealthy hyperuniform correlations from transport data. As a proof of concept, we study spinless fermions in a one-dimensional tight-binding chain connected to clean semi-infinite leads, with on-site disorder generated by imposing a stealthy spectrum $S(k)=\Theta(|k|-K)$, where $K=2\pi\chi$. 
The energy-dependent transmittance is computed using a recursive Green's function method and compared with target spectra through a misfit function defined over an energy window. We show that the position of the sharp drop separating high- and low-transmittance regions is strongly controlled by $\chi$, while the disorder strength $W$ mainly affects the absolute magnitude of the transmittance. As a result, the misfit function displays a clear minimum close to the target stealthy parameter. 
Our results demonstrate that transmittance spectra can serve as fingerprints of stealthy hyperuniform disorder, providing a practical route to infer correlated-disorder parameters from transport measurements.
\end{abstract}
\maketitle

\section{Introduction}
\label{sec:intro}

Many-particle hyperuniform disordered systems constitute a special class of correlated structures in which long-wavelength density fluctuations are strongly suppressed compared with ordinary disordered media~\cite{Torquato2003,Torquato2018}. 
In reciprocal space, this suppression is encoded in the small-$k$ behavior of the structure factor $S(\mathbf{k})$, which vanishes as $|\mathbf{k}| \to 0$~\cite{Torquato2003,Zachary2011,Torquato2018}. 
Of particular interest in photonics are so-called “stealthy” hyperuniform structures, for which $S(\mathbf{k})$ vanishes identically over a finite interval of wavevectors, $S(\mathbf{k})=0$ for $0<|\mathbf{k}|< K$~\cite{Uche2004,Batten2008,Torquato2015,Florescu2009}, forbidding single scattering from density fluctuations with wavelengths exceeding $2\pi/ K$~\cite{Leseur2016,Kim2023, Kim2024}. 
For instance, Ref.\,\onlinecite{klatt_wave_2022} showed that two-dimensional disordered network solids exhibit a stable photonic band gap only when they are sufficiently stealthy and hyperuniform. 
Weaker forms of structural order, including non-stealthy hyperuniformity, instead produce progressively shallower pseudogaps that disappear in the thermodynamic limit.
The size of the stealthy region is commonly quantified by the stealthiness parameter $\chi$, which, in the thermodynamic limit, is given by $\chi=v_1(K)/[2d\rho(2\pi)^d]$, where $\rho$ is the particle density and $v_1(K)$ is the volume of a $d$-dimensional sphere of radius $K$~\cite{Torquato2015,Klatt2026}. 
That is, the stealthiness parameter $\chi$ quantifies the fraction of constrained Fourier modes and sets the scale over which correlated disorder modifies wave propagation~\cite{Leseur2016,FroufePerez2017,Kim2023,Kim2024,kim2026}.

The significance of $\chi$ is particularly pronounced in wave and quantum transport and localization problems. 
In standard one-dimensional disordered systems, any arbitrarily weak uncorrelated disorder localizes all single-particle states in the thermodynamic limit~\cite{Anderson1958,Abrahams1979,Lee1985,Sheng2006}. 
In contrast, stealthy hyperuniform correlations suppress specific scattering processes and may produce very large localization lengths over broad energy windows~\cite{Izrailev1999,Izrailev2012,Kim2023,Klatt2026,Vanoni2026}. 
For electronic transport in general, this effect can be understood from the momentum-space structure of the random potential: the absence of Fourier components for $|{\bf k}| < K$ eliminates direct backscattering processes whenever the relevant momentum transfer lies inside the stealthy region~\cite{Izrailev1999,Izrailev2012,Vanoni2026}. 
As $\chi$ is increased, 
a larger fraction of scattering processes is eliminated, driving the system into transport regimes that behave as effectively ballistic or delocalized over accessible length scales despite the presence of disorder~\cite{Vanoni2026}.
Consequently, the stealthiness parameter $\chi$ acts as a direct knob controlling the quantum transport response~\cite{Kim2023,Kim2024,Klatt2026,Vanoni2026}.

Despite its importance, the practical determination of $\chi$ is not straightforward. 
In realistic samples, the boundary of the stealthy region may be blurred by finite-size effects, and experimental or numerical noise can obscure whether the low-$k$ spectral weight truly vanishes~\cite{Torquato2018,Kim2018,Chieco2017}.
Moreover, different real-space configurations may appear similarly disordered at short length scales while differing substantially in their long-wavelength spectral constraints~\cite{Torquato2003,Torquato2018} (see also Fig.~\ref{fig:chi_arrow}). 
These difficulties are amplified in electronic or photonic transport experiments, where the disorder characteristics are often not directly accessible.
Instead, experimental probing relies on response functions, such as conductance or transmittance, rather than the structure factor itself~\cite{Man2013,Leseur2016,FroufePerez2017,Meek2026}.

This motivates an alternative route: 
rather than attempting to extract $\chi$ from direct structural imaging, one can infer it directly from the transport fingerprints produced by that disorder. 
Such a strategy connects naturally to quantum inverse-problem approaches, in which the underlying Hamiltonian or its disorder parameters are extracted from experimentally accessible transport observables. 
In particular, recent studies~\cite{Mukim2020,Mukim2022a,Mukim2022b,Duarte2021,Duarte2024,Mukim2025,Macedo2026} have established that energy-dependent conductance spectra encode sufficient information to infer disorder characteristics through the minimization of a misfit function that compares a target conductance curve with configurationally averaged model predictions. 
The central insight is that the complex, fluctuating energy dependence of the conductance, often viewed as hard to extract physical information from, can serve as an information-rich signature of the underlying disorder potential. 
While successfully applied to uncorrelated systems, extending this inverse protocol to non-trivial correlated media provides a robust framework to quantitatively recover the stealthiness parameter from transport data alone.

For stealthy hyperuniform disorder, the parameter $\chi$ leaves a distinct feature on the transmission spectrum because it changes the set of allowed scattering wavevectors and, consequently, its energy dependence~\cite{Leseur2016,FroufePerez2017,Kim2023,Klatt2026,Vanoni2026}.
This may provide a direct route toward applications in photonics, where frequency-resolved transmission spectra are accessible. 
Indeed, experiments on hyperuniform disordered photonic structures have revealed isotropic photonic band gaps and distinct transport regimes through their transmission response~\cite{Man2013,Aubry2020,Scheffold2022,Siedentop2024,Vynck2023,barsukova2026}.
That is, such spectra are natural input observables for an inversion protocol aimed at determining the stealthiness parameter, or more generally, the long-wavelength structural correlations.

In view of this, 
in this work we demonstrate that the stealthy hyperuniform parameter can be quantitatively recovered directly
from conductance data. 
As a proof of concept, we consider a one-dimensional tight-binding chain connected to clean semi-infinite leads, with on-site disorder generated in momentum space by imposing a stealthy spectrum 
$S(\mathbf{k})=\Theta(|\mathbf{k}|-K)$~\cite{Vanoni2026,karcher2024}. 
The disorder strength $W$ controls the amplitude of the random potential, while the stealthiness parameter $\chi$ controls the size of the excluded low-$k$ region. 
Despite consisting of a one-dimensional setup, this model displays a non-trivial sequence of transitions in the scaling of the localization length $\xi$ with $W$~\cite{Vanoni2026}, making it a natural setting for probing the effectiveness of the inverse determination method for correlated disordered systems.
We compute the energy-resolved transmission via a recursive Green's function approach \cite{Thouless1981,MacKinnon1985,Lewenkopf2013} and construct an inversion protocol design to identify the parameter pair $(W,\chi)$ whose ensemble-averaged conductance spectrum best reproduces a target spectrum. 
This minimal condensed-matter platform allows us to isolate the role of stealthy correlations and establish whether transport measurements can recover $\chi$ without prior knowledge of the microscopic disorder configuration, spanning from the ballistic to the deep localized regime.

The remainder of the paper is organized as follows. In Sec.~\ref{Sec:model}, we introduce the tight-binding model with stealthy hyperuniform disorder and describe the methods used in our analysis. In Sec.~\ref{Sec:results}, we present and discuss our numerical results. Finally, in Sec.~\ref{Sec:Concl}, we summarize our main conclusions.

\section{Model and methodology}
\label{Sec:model}

\subsection{One-dimensional tight-binding model}

We consider spinless fermions on a one-dimensional tight-binding chain (denoted by $C$) of length $N$ connected to two semi-infinite clean leads (denoted by $L$ and $R$). 
The system Hamiltonian is written as $H= H_{C} + H_{\rm coupling} + H_{L} + H_R$.
The tight-binding chain Hamiltonian reads
\begin{equation}
H_{\rm C} = 
- t \sum_{j=1}^{N-1} \left( c_{j+1}^\dagger c_j + c_j^\dagger c_{j+1} \right) 
+ \sum_{j=1}^{N} w_j c_j^\dagger c_j ~,
\label{eq:central_hamiltonian}
\end{equation}
where $c_j^\dagger$ and $c_j$ are the creation and annihilation operators at site $j$, respectively, and $t$ is the nearest-neighbor hopping amplitude inside the chain. The onsite energy $w_j$ defines the disordered potential. 
The lead Hamiltonians are given by
\begin{align}
    H_L = & -t_L \sum_{j=-\infty}^0  \left(c_{j+1}^\dagger c_j + {\rm H.c.}\right),
\nonumber\\
    H_R = & -t_R \sum_{j=N+1}^\infty  \left(c_{j+1}^\dagger c_j + {\rm H.c.}\right),
\label{eq:H_leads}
\end{align}
whereas the coupling term is 
\begin{equation}
 H_{\rm coupling} = -t_L\left(c_1^\dagger c_0 + {\rm H.c.}\right) - t_R \left(c_{N+1}^\dagger c^{}_N + {\rm H.c.}\right).
\end{equation}   
Hereafter, we set $t_{L}=t_{R}=t$, taking the energy scale as a function of $t$, and define the lattice spacing to unity.

\begin{figure*}
    \centering
    \includegraphics[width=\textwidth]{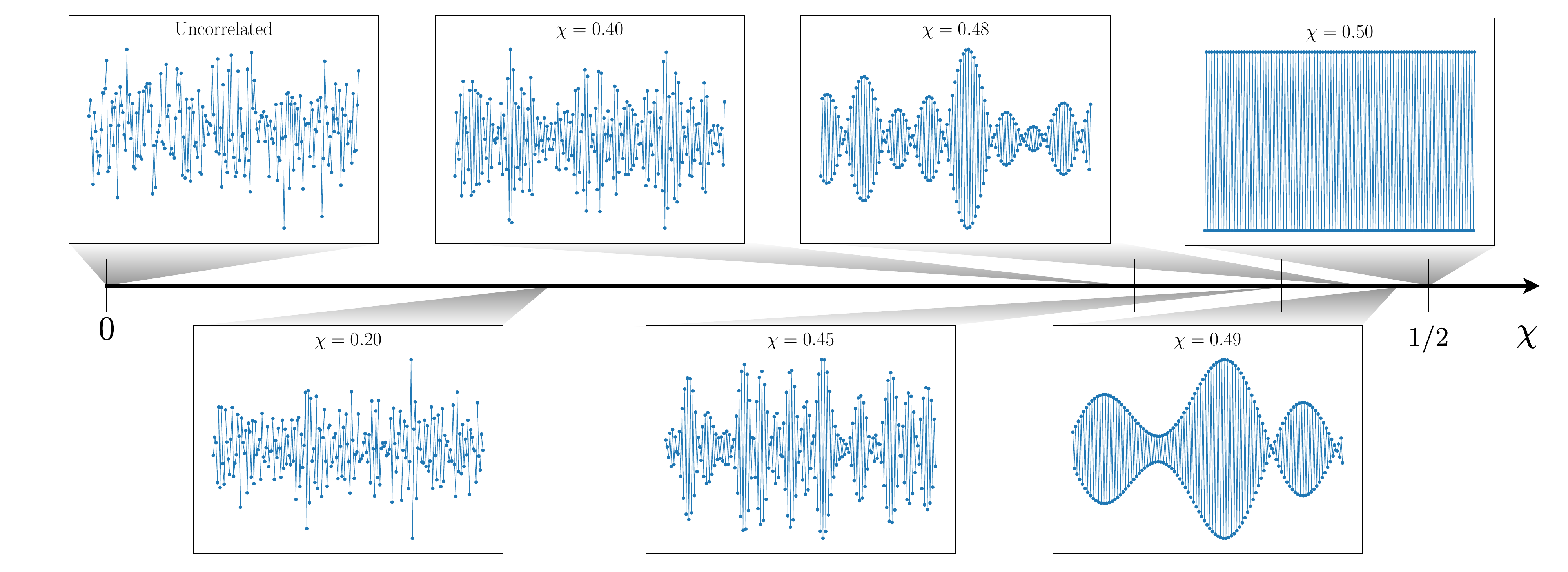}
    \caption{Real space potential $w_j$ for different values of $\chi$, ranging from the uncorrelated disorder case ($\chi =0$) on the left to the absence of disorder ($\chi = 1/2$) on the right. Correlations are readily observed at large $\chi$ values ($\chi = 0.4$ and above) but are not apparent at intermediate values ($\chi = 0.2$).}
    \label{fig:chi_arrow}
\end{figure*}

A crucial ingredient of the model is the construction of the stealthy hyperuniform disorder potential $w_j$. As described in Refs.\,\onlinecite{Vanoni2026,karcher2024}, we start from the Fourier representation of a real-space disordered potential
\begin{equation}
w_j
=
\frac{1}{\sqrt{N}}
\sum_{n}
\widetilde{w}_{k_n}
e^{i k_n j},
\label{eq:inverse_fourier}
\end{equation}
where $k_n = 2\pi n / N$ are the wavevectors, and $n=-N/2, \dots, N/2 -1$.
Since the purpose of the stealthy construction is to suppress long-wavelength fluctuations in the disorder, we impose a cutoff wavevector $K$ and remove the Fourier components with $|k|<K$. In the present work, we parametrize this cutoff as
\begin{equation}
K
=
2\pi\chi,
\label{eq:k0_chi}
\end{equation}
where $\chi$ is a dimensionless parameter controlling the size of the excluded region in momentum space.

The spectral function used to define the disorder is
\begin{equation}
S(k)
=
\Theta(|k|-K),
\label{eq:stealthy_spectrum}
\end{equation}
where $\Theta(x)$ is the Heaviside step function.
The Fourier amplitudes are generated as random variables,
\begin{equation}
\widetilde{w}_{k}
=
W\sqrt{S(k)}\,\eta_k,
\label{eq:fourier_amplitude}
\end{equation}
where $W$ controls the overall disorder strength and $\eta_k$ is a complex Gaussian random variable with zero mean. 
Since the physical potential $w_j$ must be real, the Fourier amplitudes must satisfy 
$
\widetilde{w}_{-k}
=
\widetilde{w}_{k}^{*}.
$
After the random Fourier amplitudes have been generated, the onsite disorder is obtained by the inverse Fourier transform in Eq.\,\eqref{eq:inverse_fourier}. Figure \ref{fig:chi_arrow} displays a visual representation of the real space potential across $\chi$-values.

This construction stands in contrast to uncorrelated Anderson disorder, where the potential is chosen independently at each site without restrictions on its Fourier components. 
Conversely, in the stealthy hyperuniform case, the absence of small-$k$ components produces spatial correlations in real space. 
The parameter $\chi$ controls the strength of this constraint: increasing $\chi$ widens the interval of suppressed wavevectors, thereby enhancing the hyperuniform character of the potential. 
In the limit $\chi \to 0.5$, the structure factor $S(k)$ vanishes, eliminating the disorder.  

\subsection{Recursive Green's-function method}

We compute the transport properties of the disordered chain using the recursive Green's function (RGF) method. We refer the reader to App.\,\ref{app:rec_green_1D} for a more detailed discussion. 
The key advantage of the approach is that its computational cost scales linearly with $N$, the number of sites in the central region, in contrast to the $O(N^3)$ scaling of direct diagonalization.
The concept of this recursive approach was originally developed by Thouless and Kirkpatrick \cite{Thouless1981} to calculate the linear electronic conductance of 1D atomic chains in the presence of on-site disorder. 
The method was later generalized to two-dimensional systems in the ``slice'' formulation, which is currently the most widely used version~\cite{MacKinnon1985}. 
Since then, various extensions of the algorithm have been developed to treat three-dimensional and multi-probe systems with arbitrary geometries (see, for instance, Ref.\,\onlinecite{Lima2018}).

In the site-representation, the retarded Green's function of the full system is defined as
\begin{equation}
{\bf G}^r(E) = \left( E+i\eta-{\bf H} \right)^{-1}.
\label{eq:retarded_green_function}
\end{equation}
where $\eta$ is an infinitesimal positive number. 
The partitioning of the Hamiltonian is key to solving the problem, as the semi-infinite leads do not need to be treated explicitly. 
Instead, the effect of each lead is incorporated through a self-energy acting on the boundary site of the central region.
For a semi-infinite one-dimensional lead, the surface Green's function can be obtained analytically \cite{Lewenkopf2013}. 

The coupling of the central chain to lead $\alpha=L,R$ produces the retarded self-energies
\begin{equation}
\Sigma_{\alpha}^r(E)
=
\tau_{\alpha}^2 g_{\alpha}^r(E),
\label{eq:self_energies}
\end{equation}
where $g_{\alpha}^r(E)$ is given by
\begin{equation}
g_\alpha^r(E)
=
\frac{
E+i\eta-\varepsilon_\alpha
-
\sqrt{(E+i\eta-\varepsilon_\alpha)^2-4t_\alpha^2}
}
{2t_\alpha^2}.
\label{eq:surface_gf}
\end{equation}
In line with Eq.\,\eqref{eq:H_leads}, $\varepsilon_L = \varepsilon_R = 0$.
The branch of the square root is chosen such that $\operatorname{Im} g_\alpha^r(E) \leq 0$, which ensures the retarded character of the surface Green's function and corresponds to outgoing-wave boundary conditions in the lead.

The real part of $\Sigma_\alpha^r$ shifts the energy of the boundary site, while the imaginary part describes the escape of particles from the central region into lead $\alpha$. 
The corresponding level-broadening functions are
\begin{equation}
\Gamma_\alpha(E)
=
i
\left[
\Sigma_\alpha^r(E)-\Sigma_\alpha^a(E)
\right]
=
-2\,\operatorname{Im}\Sigma_\alpha^r(E).
\label{eq:gamma}
\end{equation}

By computing the full open-system Green's function and using the above elements, one obtains the transmission from the Caroli-Meir-Wingreen formula \cite{Meir1992, Datta1995}, which for 1D systems simplifies to
\begin{equation}
T(E)
=
\Gamma_L(E)\Gamma_R(E)
\left|
G_{0,L+1}^{r}(E)
\right|^2.
\label{eq:transmission}
\end{equation}
Within the linear-response Landauer-Büttiker framework at zero temperature~\cite{Datta1995}, the two-terminal conductance for spinless fermions is directly proportional to the transmission function, $G(E) = (e^2/h) T(E)$.
Consequently, throughout this work, we formulate the inversion protocol directly in terms of the transmission spectrum.

\subsection{Inverse determination from conductance spectra}
\label{sec:inverse}

We now describe the inverse protocol used to extract the stealthiness and disorder parameters from transport data~\cite{Mukim2020,Mukim2022a,Mukim2022b}. 
The central idea is to regard the energy-dependent transmission as a fingerprint of the disorder ensemble. 
For a given pair of parameters $\Omega=(W,\chi)$, we generate independent realizations of the stealthy potential, compute the corresponding transmission spectra, and construct the configurational average
\begin{equation}
\mathcal{T}(E;\Omega)
=
\frac{1}{N_{\rm conf}}
\sum_{i=1}^{N_{\rm conf}}
T_i(E;\Omega),
\label{eq:average_transmission}
\end{equation}
where $N_{\rm conf}$ is the number of disorder realizations and $T_i(E;\Omega)$ is the transmittance of the $i$-th realization. 

The target spectrum, denoted by $\mathcal{T}_{\rm tar}(E)$, is assumed to be known within an energy interval $[E_-,E_+]$. In numerical tests, this target can be generated from a reference set of parameters $\Omega_{\rm tar}=(W_{\rm tar},\chi_{\rm tar})$, which is then treated as unknown in the inversion procedure. 
For each trial parameter set $\Omega=(W,\chi)$, we compare $\mathcal{T}_{\rm tar}(E)$ with the averaged spectrum $\mathcal{T}(E;\Omega)$ by defining the misfit function
\begin{equation}
{\cal F}(\Omega) = 
\frac{1}{E_+-E_-}
\int_{E_-}^{E_+}
dE ~
\left[
\mathcal{T}_{\rm tar}(E) -
\mathcal{T}(E;\Omega)
\right]^2 .
\label{eq:misfit_continuum}
\end{equation}
The inverse estimate is then obtained from the minimum of this function at
$\Omega_{\rm inv} =
(W_{\rm inv},\chi_{\rm inv})$.
A well-defined minimum close to $\Omega_{\rm tar}$ indicates that the conductance spectrum contains enough information to recover both the disorder strength and the stealthy parameter.

Unless otherwise stated, the target transmittance spectra are obtained by averaging over $1000$ independent disorder configurations, whereas for each trial parameter set used, the inversion is performed over $200$ realizations to optimize computational efficiency while preserving numerical stability.

In an experimental setup, the target spectrum $\mathcal{T}_{\rm tar}(E)$ is replaced by the experimental data, whose parameters are not known. The procedure described here allows one to retrieve the unknown parameters $\Omega_{\rm tar}$ of the experiment by minimizing the misfit function in Eq.\,\eqref{eq:misfit_continuum}.

\begin{figure}[t]
    \centering
    \includegraphics[width=\columnwidth]{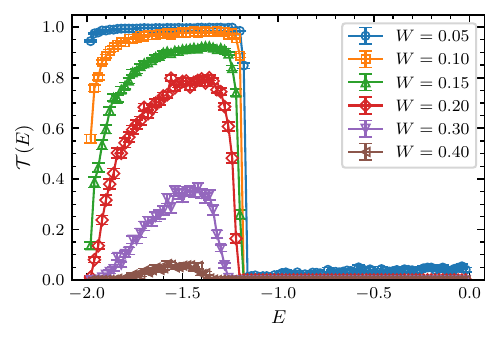}
    \caption{ Energy-resolved transmittance $\mathcal{T}(E)$ for a chain with $N=10^4$ sites, fixed stealthy parameter $\chi=0.30$, and different disorder strengths $W$. Increasing $W$ suppresses the absolute value of the transmittance inside the high-transmittance window, while the position of the sharp drop remains approximately unchanged.}
    \label{fig:TvsE_W}
\end{figure}

\section{Results}
\label{Sec:results}

We begin by examining the dependence of the energy-resolved ensemble-averaged transmittance $\mathcal{T}(E)$ on both the disorder strength $W$ and the stealthiness parameter $\chi$. Unless noted otherwise, all calculations are performed for chains of length $N=10^4$ sites. Figure~\ref{fig:TvsE_W} shows $\mathcal{T}(E)$ for a representative stealthiness value of $\chi=0.30$ and different values of the disorder strength $W$. For weak disorder, the system displays a broad energy window with large transmittance, bounded by a sharp suppression at a characteristic energy $E_c$. For the system size considered, the location of this drop is only weakly affected by changing $W$. 

\begin{figure}[t]
    \centering
    \includegraphics[width=\columnwidth]{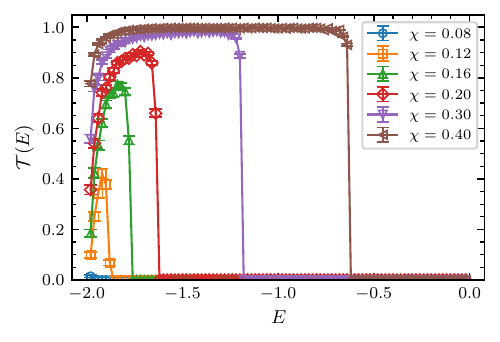}
    \caption{Energy-resolved transmittance $\mathcal{T}(E)$ for a chain with $N=10^4$ sites, fixed disorder strength $W=0.10$, and different stealthy parameters $\chi$. Increasing $\chi$ shifts the sharp drop in $\mathcal{T}(E)$ to higher energies and enlarges the high-transmittance window.}
    \label{fig:TvsE_chi}
\end{figure}

This interpretation is directly supported by comparison with perturbation theory in the weak-disorder limit $W/t \ll 1$\,\cite{Vanoni2026}, 
which gives
$E_c=-2\cos(\pi\chi)$.
For $\chi=0.30$, this analytical expression yields $E_c=-1.17557$. 
Our numerical estimates extracted from Fig.\,\ref{fig:TvsE_W} give $E_c=-1.18(1)$, $-1.20(1)$, and $-1.22(1)$ for $W=0.05$, $0.10$, and $0.20$, respectively. 
The numerical thresholds thus remain close to the perturbative value, even for $W=0.20$, which is no longer in the strictly perturbative regime. 
This quantitative agreement confirms that the boundary of the high-transmittance region is primarily set by the stealthy constraint rather than by the disorder amplitude.

Indeed, the disorder amplitude plays a secondary role in shifting $E_c$ and in reducing the overall magnitude of $\mathcal{T}(E)$ within the high-transmittance window. While the curves for small $W$ remain close to the ballistic limit over a broad energy range, stronger disorder strengths produce a substantial suppression of the transmittance prior to the threshold drop. This interplay plays an important role in the inverse problem. On one hand, the persistence of the drop at approximately the same energy indicates that the spectral signature associated with $\chi$ is robust against moderate changes in $W$. On the other hand, for larger $W$, the suppression of $\mathcal{T}(E)$ makes the jump less pronounced, which can reduce the sensitivity of the misfit function $\mathcal{F}(\Omega)$ and, in turn, make the inversion less precise.

A complementary behavior emerges when varying the stealthiness parameter $\chi$ at fixed disorder strength. Figure~\ref{fig:TvsE_chi} shows $\mathcal{T}(E)$ for $W=0.10$ and several values of $\chi$. Decreasing $\chi$ reduces the size of the stealthy window, thereby narrowing the transparent energy domain and shifting the onset of strong scattering toward lower energies. This behavior is consistent with the perturbation theory $E_c$ threshold discussed above. Within our energy resolution, the perturbative prediction agrees well with the numerical location of the drop for $\chi\gtrsim 0.16$. Therefore, while reducing $\chi$ suppresses the overall transmittance, similarly to the effect of increasing $W$, it also shifts the position of the threshold drop $E_c$ in a systematic way.

\begin{figure}[t]
    \centering
    \includegraphics[width=\columnwidth]{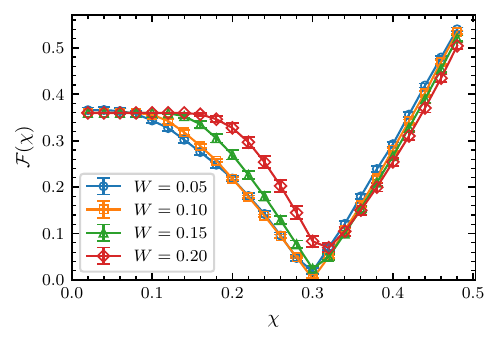}
    \caption{Misfit function $\mathcal{F}$ as a function of the trial stealthy parameter $\chi$ for a target spectrum generated with $W_{\rm tar}=0.10$ and $\chi_{\rm tar}=0.30$. 
    Distinct curves represent different trial values of $W$. 
    The minimum remains close to the target value $\chi_{\rm tar}=0.30$, displaying only small shifts when the trial disorder strength differs from $W_{\rm tar}$. Here, we analyzed a chain with $N=10^4$ sites.}
    \label{fig:misfitW01chi03}
\end{figure}

This energetic shift is highly advantageous for the inverse protocol: because $E_c(\chi)$ varies systematically with stealthiness, distinct values of $\chi$ yield clearly distinguishable conductance profiles, generating a strong variation of the cost function $\mathcal{F}(\Omega)$ near the target parameter value. Nevertheless, Fig.\,\ref{fig:TvsE_chi} also shows a key practical limitation. For small $\chi$, the states become localized over most of the band, causing $\mathcal{T}(E)\to0$ across the majority of the spectrum. In this regime, different spectra may become difficult to distinguish if the comparison is dominated by regions where $\mathcal{T}(E)\approx0$. Therefore, the integration window $[E_-,E_+]$ in Eq.\,\eqref{eq:misfit_continuum} must be judiciously selected to encompass the threshold region where $\mathcal{T}(E)$ remains strongly sensitive to the stealthy constraint.

\begin{figure*}[t]
    \centering
    \includegraphics[width=\textwidth]{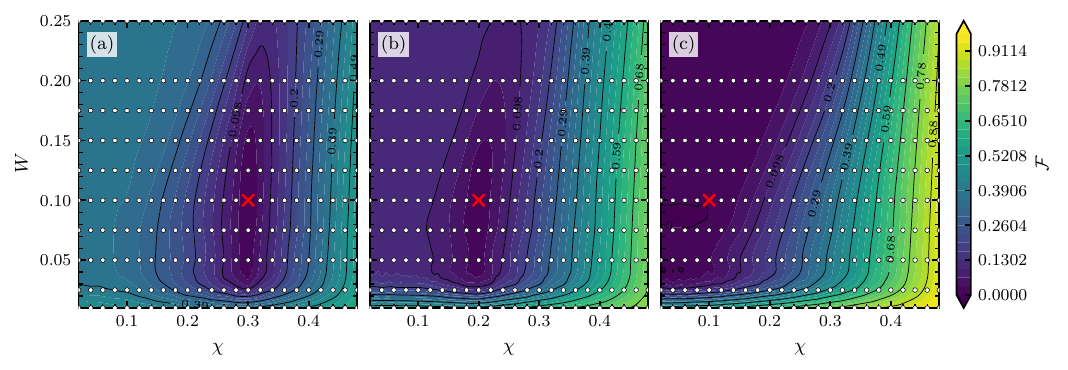}
\caption{
(a) Contour plot of the misfit function $\mathcal{F}(\chi,W)$ for a target spectrum generated with $W_{\rm tar}=0.10$ and $\chi_{\rm tar}=0.30$. The red cross indicates the target parameters. The minimum is well defined and occurs close to the target point (red cross), showing that the inverse protocol can recover both $W$ and $\chi$ when the transmittance spectrum displays a clear stealthy-induced jump. (b) Contour plot of the misfit function for a target spectrum generated with $W_{\rm tar}=0.10$ and $\chi_{\rm tar}=0.20$. (c) Contour plot of the misfit function for a target spectrum generated with $W_{\rm tar}=0.10$ and $\chi_{\rm tar}=0.10$. The red crosses determine the target values.}
\label{fig:misfit}
\end{figure*}

We now turn to the parameter reconstruction protocol directly. Figure~\ref{fig:misfitW01chi03} shows the misfit function $\mathcal{F}$ as a function of the trial stealthiness parameter $\chi$ for a target spectrum generated with $W_{\text{tar}} = 0.10$ and $\chi_{\text{tar}} = 0.30$. 
Each curve corresponds to a different trial disorder strength $W$. 
Even when $W$ deviates from $W_{\text{tar}}$, the minimum of $\mathcal{F}(\chi)$ remains sharply pinned near the true value $\chi_{\text{tar}} = 0.30$, exhibiting only minor shifts as $W$ is varied. 
These small deviations reflect a partial compensation mechanism between the two parameters: changing $W$ modifies the overall magnitude of the transmittance, while changing $\chi$ shifts the energy boundary $E_c$ of the transparency window. 
The fact that the minima remain near the target value of $\chi$ indicates that the dominant contribution to the misfit function is associated with the position of the sharp jump in $\mathcal{T}(E)$. 
This is consistent with the behavior observed in Figs.~\ref{fig:TvsE_W} and~\ref{fig:TvsE_chi}, in which the jump is weakly affected by $W$ but strongly displaced by $\chi$.

To assess the joint recovery of both parameters, we map the full cost-function landscape $\mathcal{F}(W, \chi)$ across the two-dimensional parameter space for the target set $W_{\text{tar}} = 0.10$ and $\chi_{\text{tar}} = 0.30$.
The resulting contour plot is shown in Fig.\,\ref{fig:misfit}\,(a). 
As expected, the landscape exhibits a well-defined minimum located very close to the target point. 
Therefore, whenever the stealthy parameter is sufficiently large to produce a well-resolved high-transmittance window, the conductance-based inverse protocol yields a non-degenerate solution. 
This indicates that the method can simultaneously recover the disorder strength and the stealthy parameter.

We next evaluate the performance of the inversion protocol for a smaller stealthiness parameter. 
Figure~\ref{fig:misfit}\,(b) shows the contour plot of the misfit function $\mathcal{F}(W, \chi)$ for a target spectrum generated with $W_{\rm tar}=0.10$ and $\chi_{\rm tar}=0.20$.
The minimum remains close to the target point, indicating that the conductance-based inversion remains reliable even as stealthiness decreases.
Compared with the case $\chi_{\rm tar}=0.30$, however, the minimum becomes noticeably shallower.
This behavior is consistent with the transmittance spectra discussed above: reducing $\chi$ narrows the transparent energy window and suppresses the overall dynamic range of $\mathcal{T}(E)$.
Consequently, different trial parameter pairs yield smaller spectral differences, leading to reduced sensitivity in $\mathcal{F}(W, \chi)$. 
Nevertheless, despite this flatter landscape, the global minimum remains well resolved, confirming the robustness of the reconstruction protocol for moderate stealthiness.

Finally, we analyze the breakdown limit of the inversion protocol for an even smaller value of $\chi$. 
Figure~\ref{fig:misfit}\,(c) shows the contour plot of the misfit function for a target spectrum generated with $W_{\rm tar}=0.10$ and $\chi_{\rm tar}=0.10$. 
In stark contrast to the cases with larger $\chi_{\rm tar}$, the minimum of $\mathcal{F}(W, \chi)$ is not sharply localized around the target point. 
Instead, the low-cost region forms an elongated, shallow valley, reflecting a strong loss of parameter sensitivity, particularly for small trial values of $\chi$.
This behavior reflects the fact that, for this combination of $W$ and $\chi$, the localization length $\xi$ becomes smaller than the system size $N = 10^4$ across nearly the entire energy spectrum.  
As a result, the transmittance is strongly suppressed and the sharp threshold step  $\mathcal{T}(E_c)$ is smoothed out into a featureless decay. Since the inversion protocol relies on spectral features of the transmittance, the absence of a well-defined drop introduces parameter degeneracy between $W$ and $\chi$.
We expect that reducing $W$ would increase the localization length, sharpen the transmittance feature, and consequently produce a deeper and more accurate minimum in the misfit function.

Taken together, these results establish that the conductance-based inversion protocol works best when the transmittance spectrum exhibits a well-defined step separating high- and low-transmittance regions. 
In this regime, the energy scale associated with the jump provides a robust signature of the stealthy parameter, leading to a clear minimum in the misfit function. 
By contrast, large disorder strengths or very small values of $\chi$ suppress the transmittance over most of the spectrum, reducing the sensitivity of the misfit function and making the inverse determination more challenging.

\begin{figure}[t]
    \centering
    \includegraphics[width=\columnwidth]{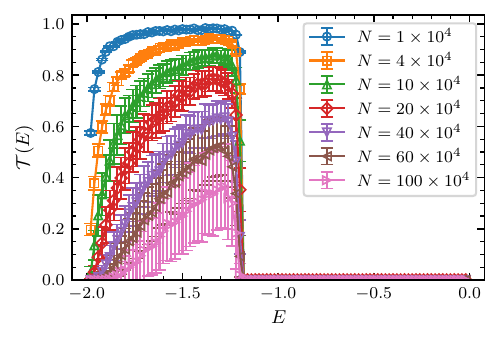}
    \caption{Energy-resolved transmittance $\mathcal{T}(E)$ for different system lengths $N$ up to $10^6$ sites, for fixed disorder strength $W=0.10$ and stealthiness parameter $\chi=0.30$.}
    \label{fig:TvsE_L}
\end{figure}

All preceding results were obtained for a fixed system size $N = 10^4$.
To verify that our results are robust against finite-size effects, we analyze the evolution of the transmittance spectrum $\mathcal{T}(E)$ as a function of the system size $N$, keeping the stealthy parameter fixed at a relatively large value.
Figure~\ref{fig:TvsE_L} shows $\mathcal{T}(E)$ for $W=0.1$ and $\chi=0.30$ across system sizes spanning two orders of magnitude, up to $N = 10^6$ sites.
The main effect of increasing $N$ is a reduction in the transmittance $\mathcal{T}(E)$ inside the high-transmittance window. 
Since one-dimensional disordered systems are expected to localize in the thermodynamic limit, this behavior is consistent with that expectation, even though the localization properties of stealthy hyperuniform disordered systems remain a matter of debate.
Despite this, the sharp drop separating the high- and low-transmittance regions remains clearly visible even for $N=10^6$. 
This indicates that, although the magnitude of the transmittance decreases with $N$, the most distinct spectral feature associated with the stealthy constraint remains robust over the entire range of sizes considered here.

\begin{figure}[t]
    \centering
    \includegraphics[width=\columnwidth]{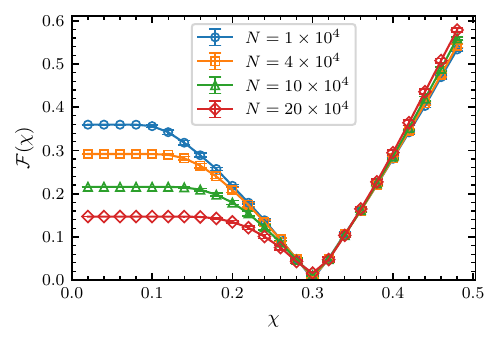}
    \caption{Misfit function $\mathcal{F}$ as a function of the trial stealthy parameter $\chi$ for a target spectrum generated with $W_{\rm tar}=0.10$ and $\chi_{\rm tar}=0.30$. Different curves correspond to different system sizes.}
    \label{fig:misfit_L}
\end{figure}

We now analyze how the system size affects the misfit function. 
Based on the transmittance spectra presented in Fig.~\ref{fig:TvsE_L}, one may expect the minimum of the misfit function to remain robust, since the sharp drop in $\mathcal{T}(E)$ persists up to $N = 10^6$.
Figure~\ref{fig:misfit_L} confirms this stability, displaying $\mathcal{F}(\chi)$ across several different system sizes for target parameters $W_{\rm tar}=0.10$ and $\chi_{\rm tar}=0.30$. 
The minimum remains clearly located around $\chi_{\rm tar}$ for all system sizes. 
The main effect of increasing $N$ is observed for trial values $\chi<\chi_{\rm tar}$, where the value of the misfit function is progressively reduced. 
This behavior is consistent with the decrease of the transmittance inside the high-transmittance window as the system size increases.

As $N \to \infty$, one eventually reaches a regime in which the system length exceeds the localization length, $N \gg \xi(E)$.
In this case, the transmittance becomes strongly suppressed 
across the entire energy band, washing out the step-like feature at $E_c$ and flattening the global minimum of the misfit function. 
This is analogous to the behavior observed in Fig.\,\ref{fig:misfit}\,(c), where the absence of a well-developed transmittance jump reduces the sensitivity of the inversion protocol.
Since a detailed 
scaling theory of localization in stealthy hyperuniform lattices lies beyond the scope of the present 
work, we contextualize our inversion protocol within the physically relevant mesoscopic regime, $N \lesssim \xi(E)$, where the stealthiness-induced transmission edge remains distinctly resolved.
Within this broad operational window, the conductance-based inverse protocol provides a highly robust and reliable scheme to reconstruct the stealthiness parameter.

Within this context, it is useful to discuss the typical system sizes realized in experiments on stealthy hyperuniform photonic structures. From this perspective, the finite-size character of the inversion protocol does not constitute a severe limitation. For instance, Ref.\,\onlinecite{Man2013} measured frequency-resolved transmission through a two-dimensional stealthy hyperuniform network with a linear size of approximately $13a$, while Ref.\,\onlinecite{Aubry2020} considered arrays containing about $200$ high-permittivity cylinders. Larger structures were achieved in Ref.\,\onlinecite{Siedentop2024}, where a three-dimensional stealthy hyperuniform network containing about $1000$ vertices was fabricated in a cubic sample. 
Even larger two-dimensional samples were recently fabricated in Ref.\,\onlinecite{barsukova2026}, which investigated stealthy-hyperuniform silicon photonic-crystal slabs with \(1596\times1596\) sites.
Ultracold atomic gases provide another promising platform for realizing correlated disorder. In quasi-one-dimensional matter-wave experiments, laser-speckle potentials extending over approximately $4$ mm—corresponding to several thousand speckle grains—have been realized~\cite{billy2008,sanchezpalencia2007}. Alternatively, programmable momentum-space lattices offer direct control over onsite energies in tight-binding models, although present implementations typically contain only a few tens of sites~\cite{meier2016,an2017}.
Although the characteristic size of these photonic and atomic platforms cannot be mapped directly onto the length $L$ of a tight-binding chain, these examples show that current transport experiments are performed in finite, mesoscopic systems. The range of system sizes investigated here therefore encompasses those currently relevant to hyperuniform-photonics experiments.

\section{Conclusions}\label{Sec:Concl}

In this work, we proposed a conductance-based inverse protocol to determine the stealthy parameter of one-dimensional hyperuniform disorder. As a proof of concept, we considered a tight-binding chain with onsite disorder generated by imposing a stealthy spectrum $S(k)=\Theta(|k|-K)$, with $K=2\pi\chi$, and computed the transmittance $\mathcal{T}(E)$ using a recursive Green's function method. We showed that the stealthy parameter $\chi$ leaves a clear fingerprint in the energy-dependent transmittance, mainly through the position of the sharp drop separating high- and low-transmittance regions. By contrast, the disorder strength $W$ primarily controls the absolute magnitude of $\mathcal{T}(E)$ inside the high-transmittance window. This separation of roles allows the misfit function to identify the target parameters with good accuracy whenever the stealthy-induced jump remains well resolved. In particular, for sufficiently large $\chi$, the misfit function develops a clear minimum close to the target values of both $W$ and $\chi$, demonstrating that transmittance spectra can be used to recover stealthy hyperuniform correlations without direct access to the microscopic disorder configuration.

Our results also clarify the limitations of the method. When the disorder strength is large, or when $\chi$ is very small, the localization length may become shorter than the system size, strongly suppressing the transmittance over most of the spectrum. In this regime, the jump in $\mathcal{T}(E)$ becomes less pronounced or may disappear, making the inverse determination less accurate. A related issue arises as the system size is increased: although we find that the characteristic jump remains visible up to $N=10^6$ sites for the parameters considered here, the transmittance inside the transparency window decreases with increasing $N$. Therefore, in the true thermodynamic limit, if $\mathcal{T}(E)$ vanishes over the whole spectrum, distinguishing different values of $\chi$ from transport data alone may become increasingly difficult.

However, this does not represent a severe limitation for realistic applications. Current experiments on hyperuniform photonic and atomic structures span a broad range of finite, mesoscopic system sizes, from a few tens of individually controlled synthetic lattice sites to photonic-crystal slabs containing thousands of elements. Although these sizes cannot be mapped directly onto the length $L$ of a one-dimensional chain, they overlap with the range explored here. In this experimentally relevant regime, the stealthy-induced spectral features remain accessible, suggesting that conductance- or transmittance-based inversion can provide a powerful methodology for determining stealthy parameters in correlated disordered media.

\begin{acknowledgments}
The authors acknowledge financial support from the Brazilian funding agencies Conselho Nacional de Desenvolvimento Cient\'\i fico e Tecnol\'ogico (CNPq), Coordena\c c\~ao de Aperfei\c coamento de Pessoal de Ensino Superior (CAPES), and Fundação de Amparo \`a Pesquisa do Estado do Rio de Janeiro (FAPERJ).
N.C.C.~acknowledges support from FAPERJ Grants No.~E-26/200.258/2023 [SEI-260003/000623/2023] and E-26/210.592/2025 [SEI-260003/004500/2025], CNPq Grants No.~313065/2021-7 and 308130/2025-1, Serrapilheira Institute Grant No.~R-2502-52037, and Alexander von Humboldt Foundation.
This publication has emanated from research conducted with the financial support of Research Ireland, Grant Number 12/RC/2278\_2, and is co-funded under the European Regional Development Fund under the AMBER award. The work of PS, ST, and CV is supported by the Army Research Office under Cooperative Agreement No. W911NF-22-2-0103.
\end{acknowledgments}

\appendix
\section{Recursive Green's functions in 1D}
\label{app:rec_green_1D}

This appendix presents the recursive Green's function (RGF) method in a nutshell for one-dimensional systems modeled by a tight-binding Hamiltonian with nearest-neighbor hopping. 
While the RGF algorithm is typically formulated for higher-dimensional systems and arbitrary hopping terms, its 1D implementation is enormously simplified and far more transparent.

The core idea is to build the system's Green's function iteratively. 
We begin by attaching the left lead to the first site ($n=1$) of the central region, keeping it temporarily disconnected from all subsequent sites ($n > 1$). 
The remaining sites of the central chain are then connected one by one. This recursive process is repeated until we reach the last site ($n = L$) of the central region, which is finally connected to the right lead.
The resulting Green's function describes the full, coupled system.

We initialize the procedure at the first site ($n=1$). The Green's function of this site, which includes the self-energy of the left lead, is given by
\begin{equation}
G_{1,1}^{L}
=
\left[
z-\varepsilon_1-\Sigma_L^r
\right]^{-1},
\label{eq:first_left_gf}
\end{equation}
and $\Sigma_L^r = \tau_L^2 g_L^r$ is the retarded self-energy of the left lead. Here, the superscript $L$ indicates that this is a “left-connected” Green's function, accounting for the subsystem built from the left up to the current site.

The remaining sites are then added sequentially. Upon adding site $n$, its local Green's function is updated via the recursion relation
\begin{equation}
G_{n,n}^{L}
=
\left[
z-\varepsilon_n
-
t^2G_{n-1,n-1}^{L}
\right]^{-1}.
\label{eq:left_recursion}
\end{equation}
This equation has a direct physical interpretation: the first two terms, $z-\varepsilon_n$, describe the isolated site $n$, while the term $t^2 G_{n-1,n-1}^{L}$ acts as an effective self-energy representing the entire portion of the chain already constructed to its left. 
Thus, each recursion step dresses the newly added site with all multiple-scattering processes occurring in the preceding chain segment and the left lead.

To compute transport properties, we also need to recursively propagate the off-diagonal Green's function that connects the surface of the left lead ($n=0$) to the current site $n$. 
We denote this left-connected off-diagonal component as $G_{0,n}^{L}$, which is updated according to
\begin{equation}
G_{0,n}^{L}
=
G_{0,n-1}^{L}
(-t)
G_{n,n}^{L},
\qquad
n=2,\dots,L,
\label{eq:offdiagonal_recursion}
\end{equation}
subject to the initial value
\begin{equation}
G_{0,1}^{L}
=
g_L^r
(-\tau_L)
G_{1,1}^{L}.
\label{eq:offdiagonal_initial}
\end{equation}
Here, $-t$ represents the hopping matrix element between neighboring sites within the central chain, while $-\tau_L$ is the coupling hopping between the left lead and the first site.

Once the recursion reaches the last site $n=L$, the right lead is attached. This connection is established by adding the right lead's self-energy to the final site. 
Equivalently, the full retarded Green's function connecting the surface of the left lead to the surface of the right lead ($n=L+1$) can be written as
\begin{equation}
G_{0,L+1}^{r}
=
\frac{
G_{0,L}^{L}
(-\tau_R)
g_R^r
}{
1-\tau_R^2 g_R^r G_{L,L}^{L}
}.
\label{eq:g0lp1}
\end{equation}
where $g_R^r$ is the uncoupled surface Green's function of the right lead and $-\tau_R$ is the coupling hopping between the last site $L$ and the right lead.

\bibliography{refs2}

\end{document}